\documentclass[11pt]{article}

\usepackage[brazil]{babel}
\usepackage{url}
\usepackage{amssymb,amsxtra} %amsthm, amsxtra,dsfont}%}%
\usepackage{amsmath,color,graphicx}

\usepackage{indentfirst}
\RequirePackage{geometry}
\usepackage[T1]{fontenc}
\usepackage[bitstream-charter]{mathdesign}

\def\na{-\kern-.4em\raise.8ex\hbox{{\tt \scriptsize a}}\ }
\def\no{-\kern-.4em\raise.8ex\hbox{{\tt \scriptsize o}}\ }
\def\ra{-\kern-.4em\raise.8ex\hbox{{\tt \tiny a}}\ }

\begin{document}

\def\chaptername{}
\def\contentsname{Sum\'{a}rio}
\def\listfigurename{Figuras}
\def\listtablename{Tabelas}
\def\abstractname{Resumo}
\def\appendixname{Ap\^{e}ndice}
\def\refname{\large Refer\^{e}ncias}
\def\bibname{Bibliografia}
\def\indexname{\'{I}ndice remissivo}
\def\figurename{\small Figura~}
\def\tablename{\small Tabela~}
\def\pagename{\small P\'{a}gina}
\def\seename{veja}
\def\alsoname{veja tamb\'em}

\setcounter{tocdepth}{3}

\clearpage
\pagenumbering{arabic}
\thispagestyle{empty}
\parskip 5pt
\vspace*{0.2cm}
\begin{center}
{\huge \bf Calores espec\'{\i}ficos dos gases ideais degenerados}\\
({\it Specific heats of degenerate ideal gases}) \\
\vspace*{0.3cm}
\vspace*{1.5cm}
{\Large \bf \it Francisco Caruso$^{1,2}$,  Vitor Oguri$^{2}$ e Felipe Silveira$^{2}$} \\*[2.em]

{{$^{1}$ Coordena\c{c}\~{a}o de F\'{\i}sica de Altas Energias\\ Centro Brasileiro de Pesquisas F\'{\i}sicas}}\\
{Rua Dr. Xavier Sigaud, 150 -- Urca, Rio de Janeiro, RJ -- 22290-180}\\*[2em]

{{$^{2}$ Departamento de F\'{\i}sica Nuclear e Altas Energias \\
Instituto de F\'{\i}sica Armando Dias Tavares\\ Universidade do Estado do Rio de Janeiro}}\\
{Rua S\~{a}o Francisco Xavier, 524 -- Maracan\~{a}, Rio de Janeiro, RJ -- 20550-900}\\*[2em]

\vfill
\end{center}

\newpage

\clearpage
\pagenumbering{arabic}
\pagestyle          {myheadings}

\vspace*{0.2cm}
\begin{center}
{\huge \bf Calores espec\'{\i}ficos dos gases ideais degenerados}\\
({\it Specific heats of degenerate ideal gases}) \\
\vspace*{0.3cm}
\vspace*{2cm}

\end{center}

\vspace*{1.5cm}

\noindent \textbf{Resumo}

 A partir de argumentos baseados no princ\'{\i}pio da incerteza de Heisenberg e no princ\'{\i}pio de exclus\~{a}o de Pauli, estimam-se os calores espec\'{\i}ficos molares dos gases ideais degenerados em baixas temperaturas, com resultados compat\'{\i}veis com o princ\'{\i}pio de Nerst-Planck (a 3\na lei da Termodin\^{a}mica). \'{E} apresentado, ainda, o fen\^{o}meno da condensa\c{c}\~{a}o de Bose-Einstein com base no  comportamento do calor espec\'{\i}fico de gases de b\'{o}sons massivos e n\~{a}o relativ\'{\i}sticos. \\
 {\bf Palavras-chave:} calor espec\'{\i}fico, gases degenerados, condensa\c{c}\~{a}o de Bose-Einstein.

\vspace*{1.5cm}
\noindent \textbf{Abstract}

From arguments based on Heisenberg's uncertainty principle and Pauli's exclusion principle, the molar specific heats of degenerate ideal gases at low temperatures are estimated, giving rise to values consistent with the Nerst-Planck Principle (third law of Thermodynamics).  The Bose-Einstein condensation phenomenon based on the behavior of specific heat of massive and non-relativistic boson gases is also presented.

\vfill

%\newpage
%\thispagestyle{empty}
%\vspace*{1mm}
%\tableofcontents

\newpage

\section{Introdu\c{c}\~{a}o}

Ap\'{o}s um longo tempo de ensino dos fen\^{o}menos t\'{e}rmicos em cursos de F\'{\i}sica, \'{e} f\'{a}cil perceber a necessidade de uma abordagem que facilite aos alunos a transi\c{c}\~{a}o da Termodin\^{a}mica para a F\'{\i}sica Estat\'{\i}stica.

A abordagem estat\'{\i}stico-probabil\'{\i}stica, tanto em seus aspectos formais como conceituais, n\~{a}o \'{e} do dom\'{\i}nio da maioria dos alunos, mesmo daqueles que j\'{a} superaram o ciclo b\'{a}sico do ensino superior. Por exemplo, muitos estudantes t\^{e}m dificuldades em compreender a explica\c{c}\~{a}o do calor espec\'{\i}fico dos gases e dos s\'{o}lidos baseada no princ\'{\i}pio da equiparti\c{c}\~{a}o de energia, pois ainda n\~{a}o sabem utilizar a distribui\c{c}\~{a}o de Maxwell-Boltzmann, ou qualquer distribui\c{c}\~{a}o de probabilidades, para o c\'{a}lculo de valores m\'{e}dios.\footnote{~Essa mesma dificuldade, por mais estranha que pare\c{c}a, \'{e} encontrada na Mec\^{a}nica Qu\^{a}ntica, ap\'{o}s a interpreta\c{c}\~{a}o probabil\'{\i}stica de Born.}

O calor espec\'{\i}fico expressa a capacidade de uma subst\^{a}ncia absorver energia quando excitada por algum agente externo. Quanto maior o n\'{u}mero de modos pelos quais \'{e} poss\'{\i}vel essa absor\c{c}\~{a}o maior o calor espec\'{\i}fico de uma subst\^{a}ncia. Por esse motivo, o calor espec\'{\i}fico de um g\'{a}s monoat\^{o}mico \'{e} menor do que a de um g\'{a}s poliat\^{o}mico, ou de um s\'{o}lido.

Desde sua descoberta~\cite{MCKIE}, os estudos e as medidas dos calores espec\'{\i}ficos t\^{e}m contribu\'{\i}do de forma determinante para a compreens\~{a}o da estrutura da mat\'{e}ria. Por exemplo, as medidas dos calores espec\'{\i}ficos dos s\'{o}lidos por P.L.~Dulong e A.T.~Petit (1819)~\cite{DULONG} permitiram que  J.J.~Berzelius~\cite{BERZELIUS} corrigisse o peso at\^{o}mico de v\'{a}rios elementos qu\'{\i}micos ao longo do s\'{e}culo~XIX.  Esse trabalho sistem\'{a}tico do qu\'{\i}mico sueco foi fundamental para que D.~Mendeleiev~\cite{MENDELEIEV} pudesse elaborar a sua Tabela Peri\'{o}dica, em torno de 1869.

%Por outro lado, o desacordo entre as predi\c{c}\~{o}es e as medidas dos calores espec\'{\i}ficos dos s\'{o}lidos e dos gases poliat\^{o}micos exigiu a revis\~{a}o cr\'{\i}tica de v\'{a}rios conceitos da F\'{\i}sica Cl\'{a}ssica, levando a modifica\c{c}\~{o}es profundas, n\~{a}o dos fundamentos da F\'{\i}sica Estat\'{\i}stica, como se acreditava, mas da pr\'{o}pria Mec\^{a}nica Cl\'{a}ssica~\cite{CARUSO-OGURI}.

Embora para gases monoat\^{o}micos os primeiros resultados, baseados no princ\'{\i}pio da equiparti\c{c}\~{a}o da energia, tenham sido satisfat\'{o}rios, o mesmo n\~{a}o ocorreu para  l\'{\i}quidos,  s\'{o}lidos e gases de mol\'{e}culas mais complexas.  A comprova\c{c}\~{a}o de que os calores espec\'{\i}ficos variavam com a temperatura, ao contr\'{a}rio da lei de Dulong-Petit, exigiu a revis\~{a}o cr\'{\i}tica de v\'{a}rios conceitos f\'{\i}sicos, levando a modifica\c{c}\~{o}es profundas, n\~{a}o dos fundamentos da F\'{\i}sica Estat\'{\i}stica, como se acreditava, mas da pr\'{o}pria Mec\^{a}nica Cl\'{a}ssica~\cite{CARUSO-OGURI}. Foram as medidas desses calores espec\'{\i}ficos dos s\'{o}lidos a baixas temperaturas que permitiram testar a ent\~{a}o nova teoria at\^{o}mica da mat\'{e}ria.

%Al\'{e}m disso, como veremos, os calores espec\'{\i}ficos variam com a temperatura, ao contr\'{a}rio da predi\c{c}\~{a}o cl\'{a}ssica de Dulong-Petit. Foram as medidas desses calores espec\'{\i}ficos dos s\'{o}lidos a baixas temperaturas que permitiram testar a ent\~{a}o nova teoria at\^{o}mica da mat\'{e}ria, a partir dos modelos de Einstein e Debye aplicado aos cristais~\cite{CARUSO-OGURI}.

As bases da F\'{\i}sica Estat\'{\i}stica foram estabelecidas por L.~Boltzmann (1884) \cite{BOLTZ} e J.~W.~Gibbs (1901) \cite{GIBBS}, a partir do conceito cl\'{a}ssico de estado de um sistema como pontos de um cont\'{\i}nuo.
%, denominado espa\c{c}o de fase.
No entanto, a descoberta de Planck (1900)~\cite{PLANCK}, generalizada pela Mec\^{a}nica Qu\^{a}ntica \cite{DIRAC}, de que os estados de um sistema, confinado em um volume, constituem um conjunto discreto associado a um espectro discreto de energia, implica a discretiza\c{c}\~{a}o do pr\'{o}prio espa\c{c}o de fase, independentemente de qualquer conceito estat\'{\i}stico.

Do ponto de vista qu\^{a}ntico, as part\'{\i}culas de um sistema, em baixas temperaturas, tendem a se agrupar pelos estados associados aos menores valores de energia. Nesse limite h\'{a} uma maior organiza\c{c}\~{a}o e uma diminui\c{c}\~{a}o da capacidade de excita\c{c}\~{a}o do sistema, o que implica decr\'{e}scimo da entropia e do calor espec\'{\i}fico de um sistema. Esse \'{e} o conte\'{u}do da 3\na\ lei da Termodin\^{a}mica.\footnote{~A 3\na\ lei da Termodin\^{a}mica, ou lei de Nernst, estabelece que a entropia de um sistema se aproxima de um valor constante no limite $T \to 0$, o qual, segundo Planck, \'{e} nulo. Desse modo, o calor espec\'{\i}fico tamb\'{e}m \'{e} nulo quando $T=0$.}

Como \'{e} mostrado nesse artigo, a defini\c{c}\~{a}o de quantidades e par\^{a}metros caracter\'{\i}sticos de um g\'{a}s ideal, baseada em princ\'{\i}pios qu\^{a}nticos, permite estabelecer a divis\~{a}o dos gases em duas categorias: os gases de f\'{e}rmions e os gases de b\'{o}sons. A partir dessa divis\~{a}o, obt\^{e}m-se estimativas para os calores espec\'{\i}ficos dos gases ideais  em baixas temperaturas, compat\'{\i}veis com a 3\na\ lei da Termodin\^{a}mica.

 Ao final do artigo, com base no comportamento do calor espec\'{\i}fico de um g\'{a}s de b\'{o}sons massivos e n\~{a}o relativ\'{\i}sticos, o fen\^{o}meno da condensa\c{c}\~{a}o de Bose-Einstein \'{e} apresentado como uma transi\c{c}\~{a}o de fase para um estado mais ordenado, o condensado de Bose-Einstein.

% \newpage
 \section{Limites dos gases ideais }

%\paragraph*{}
Os \, gases \, moleculares\, em condi\c{c}\~{o}es\, ambientais\, t\^{e}m\, densidades que variam da ordem de $10^{-5}$~g/cm$^3$ a $10^{-3}$~g/cm$^3$, e se comportam como gases ideais para os quais a energia interna depende apenas da temperatura e o calor espec\'{\i}fico molar \'{e} constante \cite{ FERMI1, SOMMER,  ZEMANSKY, MARIO}.

Sendo $N$ o n\'{u}mero total das part\'{\i}culas (\'{a}tomos ou mol\'{e}culas) constituintes de um g\'{a}s ideal, e $G$  o {n\'{u}mero de estados} associados  \`{a}s part\'{\i}culas,\footnote{~O {estado} de uma part\'{\i}cula \'{e} qualquer condi\c{c}\~{a}o poss\'{\i}vel, caracterizada por  valores de grandezas como a posi\c{c}\~{a}o e o {\it momentum}, ou a energia, associada \`{a} part\'{\i}cula.} a raz\~{a}o ($N/G$) entre esses n\'{u}meros permite a divis\~{a}o dos gases ideais em duas classes \cite{FERMI1, Epifanov}: {n\~{a}o degenerados}  e {degenerados}.
\begin{equation} \label{degenera1}  \left\{
\begin{array}{ll}
N/G \ll 1  & \ \ \mbox{(gases n\~{a}o degenerados)} \\
\vspace*{.2cm} \\
N/G \geq 1  &  \ \ \mbox{(gases degenerados)} \\
 \end{array}
\right.
\end{equation}

Enquanto as propriedades dos gases n\~{a}o degenerados n\~{a}o dependem da natureza de suas part\'{\i}culas constituintes, os gases degenerados evidenciam a natureza qu\^{a}ntica de suas part\'{\i}culas, a qual se reflete em seu comportamento macrosc\'{o}pico.

Em geral, o comportamento degenerado manifesta-se em baixas temperaturas ou altas densidades de um g\'{a}s, quando as part\'{\i}culas t\^{e}m de competir  pelos estados acess\'{\i}veis. Em altas temperaturas ou baixas densidades, quando o n\'{u}mero de estados acess\'{\i}veis \'{e} muito maior que o n\'{u}mero de part\'{\i}culas, a competi\c{c}\~{a}o pelos estados praticamente n\~{a}o existe, e o g\'{a}s exibe comportamento n\~{a}o degenerado.

%A  partir de argumentos baseados no princ\'{\i}pio da incerteza  de Heisenberg e no princ\'{\i}pio de exclus\~{a}o de Pauli, sendo definidos quantidades e par\^{a}metros que caracterizam um sistema gasoso, permitindo a divis\~{a}o dos gases degenerados de acordo com a natureza de suas part\'{\i}culas. Ser\~{a}o obtidas estimativas para os calores espec\'{\i}ficos dos gases ideais degenerados em baixas temperaturas, compat\'{\i}veis com a 3\na\ lei da Termodin\^{a}mica,  e  apresentado o fen\^{o}meno da condensa\c{c}\~{a}o de Bose-Einstein com base no   comportamento do calor espec\'{\i}fico de um g\'{a}s  de b\'{o}sons massivos e n\~{a}o relativ\'{\i}sticos.

%Por envolver um grande n\'{u}mero de part\'{\i}culas, o  comportamento e as propriedades macrosc\'{o}picas de um g\'{a}s ideal s\~{a}o determinados por m\'{e}todos estat\'{\i}sticos. No entanto,

\section{N\'{u}mero e densidade de estados}

%\paragraph*{}
Utilizando-se  uma abordagem semicl\'{a}ssica para o c\'{a}lculo do  n\'{u}mero de estados  acess\'{\i}veis  \`{a}s part\'{\i}culas  quase livres de um g\'{a}s,\footnote{~O intervalo de tempo da intera\c{c}\~{a}o de uma  part\'{\i}cula de um g\'{a}s com outras part\'{\i}culas \'{e} bem pequeno em rela\c{c}\~{a}o ao intervalo de tempo no qual que a part\'{\i}cula se move  livremente.} contidas em um volume $V$ e com {\it momenta} que variam de zero a um dado valor  $p$, os limites que caracterizam o comportamento dos gases ideais podem ser apropriadamente quantificados.

Do ponto de vista da Mec\^{a}nica Cl\'{a}ssica \cite{GIBBS, SOMMER}, o estado de uma \'{u}nica part\'{\i}cula  \'{e} caracterizado por sua posi\c{c}\~{a}o e {\it momentum}. Assim,  a evolu\c{c}\~{a}o ao longo do tempo do estado da part\'{\i}cula pode ser representada em um espa\c{c}o  de dimens\~{a}o seis, denominado {espa\c{c}o de fase} da part\'{\i}cula, no qual cada ponto de coordenadas ($x, y, z, p_x, p_y, p_z$) caracteriza o estado din\^{a}mico dessa part\'{\i}cula.

\begin{figure}[hbtp]
\centerline{\includegraphics[width=7cm]{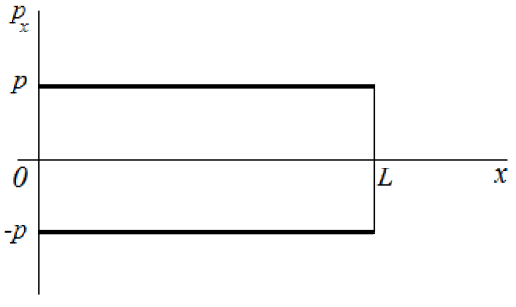}}
\caption{Plano de fase de uma part\'{\i}cula que se desloca ao longo da dire\c{c}\~{a}o $x$ com {\it momentum} entre $-p$ e $p$, confinada no intervalo $(0,L)$.}
\label{fase}
\end{figure}

Se a part\'{\i}cula move-se ao longo de uma dire\c{c}\~{a}o $x$, entre 0 e $L$, em qualquer sentido, com {\it momentum} menor que um valor $p$, sua evolu\c{c}\~{a}o \'{e} visualizada em um plano no qual cada ponto $(x, p_x)$, dentro do ret\^{a}ngulo de lados $L$ e $2 p$ (Figura~\ref{fase}), representa um poss\'{\i}vel estado acess\'{\i}vel \`{a} part\'{\i}cula. A \'{a}rea desse ret\^{a}ngulo \'{e} proporcional ao n\'{u}mero de estados acess\'{\i}veis \`{a} part\'{\i}cula.
%\footnote{~No espa\c{c}o de fases de dimens\~{a}o seis, o n\'{u}mero de  estados acess\'{\i}veis \'{e} proporcional ao  volume  $$  \int\! \int\! \int\! \int\! \int\! \int \hspace{-0.5mm} {\rm d}x {\rm d}y {\rm d}z  \, {\rm d}p_x {\rm d}p_y {\rm d}p_z .$$ }

Nesse contexto, segundo argumento originalmente proposto por O.~Sackur e H.~Tetrode~\cite{GRIMUS} em 1912,\footnote{~Esse argumento foi utilizado tamb\'{e}m por S.~Bose~\cite{BOSE} em 1924, ao deduzir a f\'{o}rmula de Planck para a radia\c{c}\~{a}o de corpo negro.} o n\'{u}mero de estados ($G$) acess\'{\i}veis a cada uma das part\'{\i}culas  de um g\'{a}s contido em um recipiente de volume $V$ pode ser expresso pela  raz\~{a}o entre o volume no espa\c{c}o de fase de dimens\~{a}o seis, associado a uma \'{u}nica part\'{\i}cula, cujo {\it momentum} varia de zero at\'{e} um  valor $p$, e um volume m\'{\i}nimo igual a $h^3$,
\begin{equation} \label{G1} \displaystyle
G = f \,  \frac{1}{h^3}  \int\! \int\! \int\! \int\! \int\! \int \hspace{-0.5mm}
{\rm d}x {\rm d}y {\rm d}z  \, {\rm d}p_x {\rm d}p_y {\rm d}p_z =
 f\, \frac{V}{h^3} \left( \frac{4\pi}{3} \ p^3 \right),
 \end{equation}
 \noindent
 sendo $h \simeq 6,\!626 \times 10^{-34}$~J$\cdot$s a constante de Planck.
 O fator $f$ depende da natureza das part\'{\i}culas,\footnote{~Para um g\'{a}s molecular ideal, $ f = 1$; para qualquer part\'{\i}cula de {\it spin} 1/2 ou part\'{\i}culas n\~{a}o massivas de {\it spin} 1 (f\'{o}ton ou o f\^{o}non) , $f = 2$. Esse fator indica a multiplicidade dos estados associados \`{a}s part\'{\i}culas.} e o termo entre par\^{e}nteses \'{e} o volume da esfera de raio $p= \sqrt{p_x^2+p_y^2+p_z^2}$ no espa\c{c}o dos {\it  momenta}.

 A exist\^{e}ncia de um volume m\'{\i}nimo no espa\c{c}o de fase pode ser justificado a partir do  princ\'{\i}pio da incerteza de Heisenberg \cite{CARUSO-OGURI,  DIRAC,  FERMI1}, o qual   estabelece uma correla\c{c}\~{a}o entre o {\it momentum} ($p_{_x}$) e a  posi\c{c}\~{a}o ($x$)   de uma part\'{\i}cula em uma dada dire\c{c}\~{a}o, tal  que o limite m\'{\i}nimo para o produto das incertezas ($\Delta_{x}, \Delta_{p_x}$)  associadas \`{a}s medidas desses pares de vari\'{a}veis \'{e} da ordem da constante de Planck.

Para as tr\^{e}s dire\c{c}\~{o}es espaciais, pode-se escrever
$$
\big( \Delta_x  \cdot \Delta_{p_x}\big)_{\mbox{\tiny min}}  \sim  h, \qquad
\big( \Delta_y  \cdot \Delta_{p_y} \big)_{\mbox{\tiny min}}  \sim h   \qquad \mbox{\small e} \qquad
\big( \Delta_z  \cdot \Delta_{p_z} \big)_{\mbox{\tiny min}}  \sim h.
$$

%\newpage
De acordo com as rela\c{c}\~{o}es entre o {\it momentum} ($p$) e a energia ($\epsilon$) de uma part\'{\i}cula livre,
\begin{equation} \label{dispersa} \displaystyle
 \left\{
\begin{array}{ll}
p = \sqrt{2m\epsilon}  & \ \ \mbox{(part\'{\i}culas n\~{a}o relativ\'{\i}sticas de massa $m$)} \\
\hspace*{1cm} & \ \ \  \\
%\hspace*{1cm} & \ \ \  \mbox{n\~{a}o relativ\'{\i}sticas),} \\
\frac{}{}\vspace*{.1cm} \\
p = \textstyle \epsilon / c &  \ \ \mbox{(part\'{\i}culas ultrarrelativ\'{\i}sticas ou} \\
\hspace*{1cm} & \ \ \   \mbox{n\~{a}o massivas com velocidade $c$: f\'{o}tons, f\^{o}nons, ...),} \\
 \end{array}
\right.
\end{equation}
\noindent pode-se expressar o n\'{u}mero de estados acess\'{\i}veis eq.~(\ref{G1}) a cada part\'{\i}cula de um g\'{a}s como fun\c{c}\~{a}o da energia
por
\begin{equation}\label{G2}
  G(\epsilon) = \left\{
\begin{array}{l}
\left( \frac{\displaystyle 4\pi f}{\displaystyle 3} \right) \ \frac{\displaystyle V}{\displaystyle h^3} \
\left( 2m\epsilon \right)^{3/2}, \\
\vspace*{.2cm} \\
\left( \frac{\displaystyle 4\pi f}{\displaystyle 3} \right) \ \frac{\displaystyle V}{\displaystyle (ch)^3} \ \epsilon^3. \\
\end{array}
\right.
\end{equation}

Para um g\'{a}s molecular nas condi\c{c}\~{o}es ambientais, o n\'{u}mero de estados acess\'{\i}veis \'{e} da ordem de  $10^{27}$ e o n\'{u}mero de mol\'{e}culas, cerca de $10^{23}$. Nessas condi\c{c}\~{o}es, o estado do g\'{a}s \'{e} n\~{a}o degenerado.

Definindo-se a {densidade de estados} $g(\epsilon)$ de uma part\'{\i}cula livre por
\begin{equation}\label{D1}
 g(\epsilon) = \frac{\mbox{d}G}{\mbox{d}\epsilon} = \left\{
\begin{array}{l}
 2\pi f V \ \left( \frac{\displaystyle 2m}{\displaystyle h^2} \right)^{3/2} \epsilon^{1/2}, \\
\vspace*{.2cm} \\
 4\pi f \ \frac{\displaystyle V}{\displaystyle (ch)^3} \ \epsilon^2,\\
\end{array}
\right.
\end{equation}
o n\'{u}mero de estados acess\'{\i}veis em qualquer intervalo de energia pode ser determinado integrando-se $g(\epsilon)$ em rela\c{c}\~{a}o \`{a} energia.

\section{Gases ideais n\~{a}o degenerados}

%\paragraph*{}
De acordo com os experimentos sobre o comportamento os gases \cite{FERMI1, SOMMER, ZEMANSKY, MARIO}, anteriores ao s\'{e}culo~XX, a energia interna ($U$) de um g\'{a}s ideal monoat\^{o}mico, em condi\c{c}\~{o}es ambientais, \`{a} temperatura $T$ e, portanto, n\~{a}o degenerado, \'{e} igual a
$$
 U  = \frac{3}{2} \,  NkT  = \frac{3}{2} \, n R T,
 $$
\noindent e o calor espec\'{\i}fico molar a volume constante ($c_{_V}$),\footnote{~O calor espec\'{\i}fico molar a volume constante \'{e} definido por
$$  c_{_V} = \frac{1}{n} \, \left( \frac{\partial U}{\partial T} \right)_V. $$}
 %= \frac{T}{n} \, \left( \frac{\partial S}{\partial T} \right)_V,$$
%em que  $S$ \'{e} a entropia do g\'{a}s.} a
%
\begin{equation}\label{calor2}
\fbox{$ \displaystyle
 c_{_V}  = \frac{3}{2} \,   R $} \qquad  \mbox{\small (gas ideal n\~{a}o degenerado monoat\^{o}mico),}
\end{equation}
\noindent sendo $k \simeq 1,\!38 \times 10^{-23}$~J/K a constante de Boltzmann, $R \simeq 8,\!3$~J$\cdot$mol$^{-1}\cdot$K$^{-1}$ a constante dos gases, $N$ o n\'{u}mero de mol\'{e}culas e $n$ o n\'{u}mero de mols.

Em baixas temperaturas, esse comportamento cl\'{a}ssico do calor espec\'{\i}fico n\~{a}o \'{e} compat\'{\i}vel com  a 3\na\ lei da Termodin\^{a}mica \cite{FERMI1, SOMMER, ZEMANSKY, MARIO}, segundo a qual o calor espec\'{\i}fico a 0~K deve-se anular,
%\footnote{~De acordo com  a   3\na\ lei da Termodin\^{a}mica, ou lei de Nernst, a entropia de um sistema se aproxima de um valor constante no limite $T \to 0$, o qual segundo Planck \'{e} nulo. Desse modo, o calor espec\'{\i}fico tamb\'{e}m \'{e} nulo quando $T=0$.}
$$ \lim_{T \to 0} c_{_V}  \ \to \ 0. $$

\section{Temperaturas cr\'{\i}ticas}

%\paragraph*{}
Uma vez que a energia m\'{e}dia por part\'{\i}cula $\bar{\epsilon} = U/N$ de um g\'{a}s n\~{a}o degenerado \'{e} cerca de
$$ \bar{\epsilon}  \simeq kT, $$
o n\'{u}mero de estados ocupados pelas part\'{\i}culas de um g\'{a}s n\~{a}o degenerado \'{e} cerca de $G(\bar{\epsilon})$ e, portanto, o crit\'{e}rio para a n\~{a}o degeneresc\^{e}ncia pode ser expresso como
\begin{equation}\label{degenera2}
 \frac{N}{G} \simeq \left\{
\begin{array}{l}
\left( \frac{\displaystyle 3}{\displaystyle 4 \pi f} \right) \
\left( \frac{\displaystyle h^2}{\displaystyle 2mk} \right)^{3/2} \
\left( \frac{\displaystyle N}{\displaystyle V} \right) \
\frac{\displaystyle 1}{\displaystyle T^{3/2}} \\
\vspace*{.2cm} \\
\left( \frac{\displaystyle 3}{\displaystyle 4\pi f} \right) \
\left( \frac{\displaystyle ch}{\displaystyle k} \right)^3 \
\left( \frac{\displaystyle N}{\displaystyle V} \right) \
\frac{\displaystyle 1}{\displaystyle T^3} \\
\end{array}
\right. \hspace*{.4cm}  \ll 1.
\end{equation}

\begin{figure}[hbtp]
\vspace*{0.2cm}
\centerline{\includegraphics[height=6.0cm]{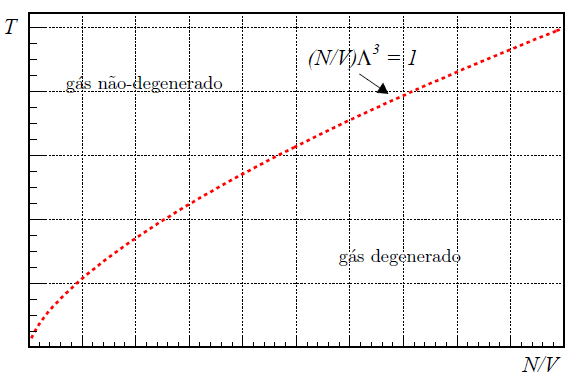}}
\vspace{-0.5cm}
\caption{Limite de  degeneresc\^{e}ncia dos  gases.}
\label{lim_class}
\end{figure}

A figura~\ref{lim_class} mostra o compromisso entre a densidade de part\'{\i}culas ($N/V$) e a temperatura  de um g\'{a}s de part\'{\i}culas massivas n\~{a}o relativ\'{\i}sticas -- a linha pontilhada\footnote{~Essa linha, de acordo com a primeira das eqs.~(\ref{degenera2}),  \'{e} definida por $(N/V) \, \Lambda^3= 1$, em que $\Lambda = \displaystyle  \left( \frac{3}{8\pi} \right)^{1/3} \! \! \frac{h}{\sqrt{2mkT}}$.}  define os valores cr\'{\i}ticos que separam as regi\~{o}es nas quais o gas \'{e}  degenerado ou n\~{a}o degenerado.

Para temperaturas e densidades abaixo dessa linha, o estado do g\'{a}s \'{e} degenerado  e a natureza qu\^{a}ntica de suas  part\'{\i}culas deve ser considerada. Para temperaturas e densidades acima da linha cr\'{\i}tica, o estado do g\'{a}s \'{e} n\~{a}o degenerado e pode ser tratado como um sistema cl\'{a}ssico.

\newpage
A Tabela~\ref{Tc} mostra os valores  cr\'{\i}ticos de temperatura  e das  densidades de alguns gases moleculares.

\renewcommand{\arraystretch}{1.4}
\begin{table}[htbp]
\caption{Valores cr\'{\i}ticos correspondentes a  densidade de part\'{\i}culas e a temperatura de alguns gases moleculares.}
\begin{center}
\begin{tabular}{l|c|c}
& \multicolumn{2}{c}{\quad valores cr\'{\i}ticos} \\
\hline
~~~~~\mbox{gases}             & ~~~$N/V$~\big(\mbox{cm}$^{-3}$\big)~~~ & ~~~$T$~(\mbox{K})~~~ \\
\hline
~~~~~~~\mbox{\tt H}$_2$            &  $ \sim 10^{19}$          &  $\sim 0,\!1$   \\
~~~~~~~\mbox{\tt He}$^4$          &   $\sim 10^{22}$          &  $\sim 3$    \\
 \hline
\end{tabular}
\end{center}
\label{Tc}
\end{table}
\renewcommand{\arraystretch}{1}

Em fun\c{c}\~{a}o da densidade de part\'{\i}culas $(N/V$), as {temperaturas cr\'{\i}ticas} ($T_c$), definidas pela condi\c{c}\~{a}o $N/G=1$, s\~{a}o dadas por
\begin{equation}\label{Tc1}
 T_c = \left\{
\begin{array}{l}
T_{_F} = \left( \frac{\displaystyle 3}{\displaystyle 4\pi f} \right)^{2/3} \
\left( \frac{\displaystyle h^2}{\displaystyle 2mk} \right) \
\left( \frac{\displaystyle N}{\displaystyle V} \right)^{2/3}, \\
\vspace*{.2cm} \\
T_{_D} = \left( \frac{\displaystyle 3}{\displaystyle 4\pi f} \right)^{1/3} \
\left( \frac{\displaystyle ch}{\displaystyle k} \right) \
\left( \frac{\displaystyle N}{\displaystyle V} \right)^{1/3},
\end{array}
\right.
\end{equation}
\noindent em que  $T_{_F}$  e $T_{_D}$  s\~{a}o usualmente denominadas {temperatura de Fermi} e {temperatura de Debye}.

Assim, o crit\'{e}rio para a n\~{a}o degeneresc\^{e}ncia de um g\'{a}s \'{e} usualmente expresso como
\begin{equation}\label{Tc2}
\fbox{$ \displaystyle
 \frac{T}{T_c} \gg 1 $} \qquad  \mbox{\small (crit\'{e}rio para a n\~{a}o degeneresc\^{e}ncia).}
\end{equation}
Se \ $T \leq T_c$, o g\'{a}s \'{e} dito degenerado.

Os gases ideais, n\~{a}o degenerados ou degenerados, s\~{a}o idealiza\c{c}\~{o}es que n\~{a}o correspondem exatamente a nenhum sistema macrosc\'{o}pico. Apesar disso, a \^{e}nfase observada em seu estudo, bastante simples, decorre do fato de que em um grande n\'{u}mero de casos alguns sistemas macrosc\'{o}picos podem ser representados por conjuntos de constituintes quase independentes, ou seja, como gases.
\renewcommand{\arraystretch}{0.5}
\begin{table}[htbp]
\caption{Temperaturas cr\'{\i}ticas de alguns sistemas t\'{\i}picos. Os tr\^{e}s primeiros exemplos s\~{a}o de sistemas de part\'{\i}culas massivas n\~{a}o relativ\'{\i}sticas e os dois \'{u}ltimos, de part\'{\i}culas n\~{a}o massivas e ultrarrelativ\'{\i}sticas, respectivamente.}
\begin{center}
\begin{tabular}{|l|c|c|} \hline
~~~~~~sistemas          &  densidade de                             & temperatura  \\
                                       &  de part\'{\i}culas~(cm$^{-3})$                      & cr\'{\i}tica (K) \\
\hline
                                       &                                               &       \\
gases                              &  $10^{19}$         &  $ < 0,\!1$  \\
moleculares                   &   &  \\
                                       &                                               &       \\
%{\tt He}$^4$ l\'{\i}quido   &  $10^{28}$~m$^{-3}$          & $2,\!7$  \\
 %                                     &                                              &       \\
           el\'{e}trons em                   &  $10^{17}$        &  $10$   \\
semicondutores             &                                              &       \\
                                        &                                               &       \\
el\'{e}trons de condu\c{c}\~{a}o     &  $10^{23}$        & $10^5$   \\
em metais                       &                                               &       \\
                              &                                              &       \\
osciladores at\^{o}micos     &  $10^{23}$        & $10^2$  \\
em cristais                      &     &       \\
                                        &                                               &       \\
%radia\c{c}\~{a}o de                  &  $10^{29}$         & $10^7$ \\
%corpo negro  (f\'{o}tons)   &  ($c \simeq  10^8$~m/s)   &       \\
 %                                     &                                               &       \\
 %                                     &                                               &       \\
estrelas do tipo              &  $10^{36}$           & $10^{10}$ \\
an\~{a} branca  (el\'{e}trons)   &      &     \\
                                        &                                               &       \\
\hline
\end{tabular}
\end{center}
\label{tcri}
\end{table}
\renewcommand{\arraystretch}{1.0}

A Tabela~\ref{tcri} mostra, de acordo com as eqs.~(\ref{Tc1}), as temperaturas cr\'{\i}ticas relativas a v\'{a}rios sistemas que se comportam  como gases  ideais. Pode-se observar, assim, que, mesmo para baixas temperaturas ($T \sim 10$~K), os gases moleculares comportam-se como gases ideais n\~{a}o degenerados; j\'{a} os el\'{e}trons de condu\c{c}\~{a}o nos metais constituem um sistema degenerado em qualquer temperatura, uma vez que a $10^5$~K n\~{a}o existe mat\'{e}ria s\'{o}lida. No caso dos el\'{e}trons relativ\'{\i}sticos provenientes dos \'{a}tomos de h\'{e}lio ionizados de uma estrela an\~{a} branca a $10^7$~K, a temperatura da estrela ainda \'{e} muito menor que  o valor  cr\'{\i}tico e, portanto, o sistema comporta-se como um g\'{a}s degenerado relativ\'{\i}stico.
Por outro lado, os el\'{e}trons em semicondutores  e os osciladores at\^{o}micos nos cristais n\~{a}o s\~{a}o degenerados nas condi\c{c}\~{o}es ambientais,  sendo degenerados apenas  em baixas temperaturas ($T \sim 10$~K).

Em termos  das temperaturas cr\'{\i}ticas, os n\'{u}meros e as densidades de estados podem ser escritos como
\begin{equation}\label{G3}
G(\epsilon) = \left\{
\begin{array}{l}
N\, \bigg(\frac{\displaystyle \epsilon}{\displaystyle kT_{_F}} \bigg)^{3/2}, \\
\vspace*{.2cm} \\
 N\, \left(\frac{\displaystyle \epsilon}{\displaystyle kT_{_D} } \right)^3,  \\
\end{array}
\right.
\qquad \Longrightarrow \qquad
g(\epsilon) = \left\{
\begin{array}{l}
 \frac{\displaystyle 3}{\displaystyle 2} \ \frac{\displaystyle N}{\displaystyle (kT_{_F})^{3/2}} \
\epsilon^{1/2}, \\
\vspace*{.2cm} \\
 3 \ \frac{\displaystyle N}{\displaystyle (kT_{_D})^3} \
\epsilon^2.\\
\end{array}
\right.
 \end{equation}

\section{Gases ideais degenerados}

%\paragraph*{}
O princ\'{\i}pio de exclus\~{a}o de Pauli \cite{CARUSO-OGURI,  DIRAC,  FERMI1} estabelece uma outra correla\c{c}\~{a}o qu\^{a}ntica tal que part\'{\i}culas id\^{e}nticas de {\it spin} semi-inteiro n\~{a}o podem compartilhar o mesmo estado. Assim, os gases ideais degenerados dividem-se em duas classes: aqueles constitu\'{\i}dos por  part\'{\i}culas de {\it spin} semi-inteiro, denominadas {f\'{e}rmions}, e aqueles constitu\'{\i}dos por part\'{\i}culas de {\it spin} inteiro, denominadas {b\'{o}sons}. Nos gases n\~{a}o degenerados, quando efetivamente n\~{a}o h\'{a} competi\c{c}\~{a}o pelos estados acess\'{\i}veis por suas part\'{\i}culas constituintes, tal distin\c{c}\~{a}o n\~{a}o \'{e} necess\'{a}ria.

Em temperaturas pr\'{o}ximas  a $0$~K, quando o calor espec\'{\i}fico e a entropia de um sistema tendem a se anular,
%\footnote{~De acordo com a 3\ra\ lei da Termodin\^{a}mica.}
as part\'{\i}culas de um g\'{a}s distribuem-se entre os estados acess\'{\i}veis tal que a energia total seja m\'{\i}nima. Diz-se que o sistema encontra-se em seu estado fundamental.

Para temperaturas finitas, mas ainda muito menores que a temperatura cr\'{\i}tica ($T \ll T_c$), uma pequena fra\c{c}\~{a}o das part\'{\i}culas do g\'{a}s s\~{a}o excitadas al\'{e}m do estado fundamental. Essas part\'{\i}culas excitadas comportam-se, praticamente, como um subsistema n\~{a}o degenerado respons\'{a}vel pelas propriedades t\'{e}rmicas do g\'{a}s.

\subsection{F\'{e}rmions  n\~{a}o relativ\'{\i}sticos fortemente degenerados}

%\paragraph*{}
Devido ao princ\'{\i}pio de Pauli, para um g\'{a}s ideal de f\'{e}rmions completamente degenerado ($T=0$~K) em seu estado fundamental, as part\'{\i}culas se acomodam nos estados acess\'{\i}veis de tal forma que  cada estado seja ocupado por apenas uma part\'{\i}cula. Nessas condi\c{c}\~{o}es, o n\'{u}mero de estados ocupados \'{e} igual ao n\'{u}mero de part\'{\i}culas.

Assim, para um g\'{a}s de f\'{e}rmions n\~{a}o relativ\'{\i}sticos completamente degenerado no estado fundamental,
$$ \frac{N}{G(\epsilon_{_F})} = 1, $$
 sendo $\epsilon_{_F}$ o maior valor de energia associado aos estados ocupados pelas part\'{\i}culas. De acordo com as eqs.~(\ref{G3}), esse valor, denominado energia de Fermi, \'{e} dado por
\begin{equation}\label{Fermi}
\fbox{$ \displaystyle \epsilon_{_F} = kT_{_F} $.}
 \end{equation}

Segundo a Tabela~\ref{tcri}, para o g\'{a}s de el\'{e}trons de condu\c{c}\~{a}o de um metal \`{a} temperatura ambiente ($T \ll T_{_F}$), a energia de Fermi \'{e} da ordem de
$$\epsilon_{_F} \simeq 10^{-18}~\mbox{J} \simeq 10~\mbox{eV},$$
e a energia dos el\'{e}trons excitados cerca de
$$\epsilon \simeq kT \simeq 25~\mbox{meV} \ll \epsilon_{_F}  \quad (T \sim 300~{\rm K}
).$$

Para temperaturas acima de $0$~K, mas ainda muito menores que a temperatura de Fermi ($T \ll T_{_F}$),\footnote{~Para os el\'{e}trons de condu\c{c}\~{a}o em um metal, essa condi\c{c}\~{a}o sempre se verifica.} algumas part\'{\i}culas s\~{a}o excitadas acima do n\'{\i}vel de Fermi. Essas part\'{\i}culas excitadas comportam-se como um subsistema n\~{a}o degenerado com energia m\'{e}dia por part\'{\i}cula da ordem de $kT \ll \epsilon_{_F}$.
 Nessas condi\c{c}\~{o}es, o n\'{u}mero de part\'{\i}culas excitadas ($N_{\mbox{\tiny exc}}$) \'{e} aproximadamente igual ao n\'{u}mero de estados acess\'{\i}veis no intervalo de energia $(\epsilon_{_F} , \epsilon_{_F} + kT)$, ou seja,
\begin{eqnarray*}
N_{\mbox{\tiny exc}} &\simeq& \int_{\epsilon_{_F}}^{\epsilon_{_F}+kT}
\frac{3}{2} \frac{N}{\epsilon_{_F}^{3/2}} \epsilon^{1/2} \, \mbox{d}\epsilon  \ \ = \ \
 \frac{N}{\epsilon_{_F}^{3/2}} \Big[ (\epsilon_{_F}+kT)^{3/2} -
\epsilon_{_F}^{3/2} \Big] \\
& \simeq   &  N \Big[ (1+kT/\epsilon_{_F})^{3/2} - 1 \Big] \ \  \simeq  \ \
\frac{3}{2} N \frac{T}{T_{_F}}.
\end{eqnarray*}

Pode-se obter a energia interna  ($U$) do g\'{a}s  adicionando-se a energia do estado fundamental ($U_{\circ}$) para $T=0$~K \`{a} energia das part\'{\i}culas excitadas ($N_{\mbox{\tiny exc}} kT$),
$$U \ \simeq \  U_{\circ} \ +  \ N_{\mbox{\tiny exc}} kT \ = \  U_{\circ} \ + \ \frac{3}{2} N k \frac{T^2}{T_{_F}},$$
\noindent
 e o calor espec\'{\i}fico molar a volume constante,  por\footnote{~Segundo Sommerfeld~\cite{SOMMER}, $\displaystyle c_{_V} \ = \  \frac{\pi^2}{2} \, R \, \frac{T}{T_{_F}} $.}
\begin{equation}\label{calor3}
\fbox{$ \displaystyle c_{_V} \ \simeq \  3 R \, \frac{T}{T_{_F}}   \quad (T \ll T_{_F}) $} \qquad  \mbox{\small (f\'{e}rmions  n\~{a}o relativ\'{\i}sticos).}
 \end{equation}

Esse comportamento, obtido por Sommerfeld em 1928, compat\'{\i}vel com a 3\na\ lei da Termodin\^{a}mica, \'{e} o esperado para a varia\c{c}\~{a}o do calor espec\'{\i}fico dos metais em temperaturas pr\'{o}ximas a $0$~K, quando as propriedades t\'{e}rmicas dos metais s\~{a}o atribu\'{\i}das ao movimento dos el\'{e}trons de condu\c{c}\~{a}o \cite{SOMMER}.

\subsection{B\'{o}sons ideais degenerados}

%\paragraph*{}
Uma vez que os b\'{o}sons de um sistema podem ser associados a estados de mesma energia, todos podem ocupar um \'{u}nico estado. Em geral, a energia do estado fundamental de um g\'{a}s de b\'{o}sons completamente degenerado ($T = 0$~K) pode ser considerada nula.

De modo an\'{a}logo aos f\'{e}rmions, para temperaturas um pouco acima de $0$~K, mas ainda muito menores que a temperatura cr\'{\i}tica ($T\ll T_{_D}$), algumas part\'{\i}culas s\~{a}o excitadas com energia da ordem de $kT$.

Dependendo da massa e do car\'{a}ter relativ\'{\i}stico, o n\'{u}mero de part\'{\i}culas excitadas ($N_{\epsilon > 0}$), bem como a energia interna ($U$) e o calor espec\'{\i}fico molar ($c_{_V}$), s\~{a}o calculados de modos distintos.

\subsubsection{B\'{o}sons n\~{a}o massivos fortemente degenerados}

%\paragraph*{}
No caso de b\'{o}sons n\~{a}o massivos, apesar de ser pequena a  fra\c{c}\~{a}o de part\'{\i}culas que abandonam o estado fundamental quando  $T \ll T_{_D}$, o n\'{u}mero de part\'{\i}culas excitadas com energia m\'{e}dia da ordem de $kT$ \'{e}  maior do que o n\'{u}mero de estados com energia at\'{e} $kT$. Assim,
$$
N_{\epsilon > 0}  = \alpha \underbrace{\int_0^{kT}
3\, \frac{N}{(kT_{_D})^3} \, \epsilon^2 \, \mbox{d}\epsilon}_{G(\epsilon >0)} =
N \alpha \, \left( \frac{T}{T_{_D}} \right)^{3},
$$
\noindent
sendo   $\alpha$  um par\^{a}metro da ordem de 10, que depende da raz\~{a}o $T/T_{_D}$.\footnote{~$\alpha = \pi^4/5 \quad  (T \ll T_{_D})$.}

Desse modo, a energia interna corresponde a
\begin{equation}\label{U3}
 \displaystyle
  U = N_{\epsilon > 0} kT =  N k \alpha \ \frac{T^4}{T_{_D}^3},
 \end{equation}
\noindent
e o calor espec\'{\i}fico molar a volume constante, a

\begin{equation}\label{calor4}
\fbox{$ \displaystyle c_{_V} =  4 R \alpha \left( \frac{T}{T_{_D}} \right)^3  \quad (T \ll T_{_D}) $} \qquad  \mbox{\small ( b\'{o}sons n\~{a}o massivos).}
 \end{equation}

A depend\^{e}ncia da temperatura ($\propto T^3$), estabelecida por P.~Debye~\cite{DEBYE}, em 1912, \'{e} compat\'{\i}vel com a 3\na\ lei da Termodin\^{a}mica, e descreve o comportamento do calor espec\'{\i}fico molar dos s\'{o}lidos  em baixas temperaturas ($ T \sim 4~\mbox{K}$) \cite{KEESOM}.\footnote{~Para os s\'{o}lidos, o calores espec\'{\i}ficos a volume e a press\~{a}o constantes s\~{a}o praticamente iguais, principalmente em baixas temperaturas.}

\subsubsection{Os s\'{o}lidos cristalinos}

%\paragraph*{}
A lei emp\'{\i}rica de Dulong e Petit, de que o valor do calor espec\'{\i}fico dos s\'{o}lidos seria uma constante independente da temperatura, apesar dos muitos desvios observados, s\'{o} come\c{c}a a declinar, realmente, no in\'{\i}cio do s\'{e}culo~XX. A partir de misturas refrigerantes, foram alcan\c{c}adas temperaturas baixas o suficiente para evidenciar a depend\^{e}ncia do calor espec\'{\i}fico com a temperatura. Verificou-se, experimentalmente \cite{KEESOM}, que, em baixas temperaturas, o calor espec\'{\i}fico de um s\'{o}lido obedecia \`{a} 3\na\ lei da Termodin\^{a}mica, variando com a temperatura segundo a chamada lei de Debye, eq.~(\ref{Debye_lei}), a seguir.

%Essa unidade b\'{a}sica \'{e} denominada c\'{e}lula unit\'{a}ria e o arranjo resultante de rede cristalina.

A representa\c{c}\~{a}o das vibra\c{c}\~{o}es at\^{o}micas nos s\'{o}lidos cristalinos\footnote{~Um s\'{o}lido cristalino \'{e} constitu\'{\i}do pela repeti\c{c}\~{a}o de uma unidade b\'{a}sica de padr\~{a}o geom\'{e}trico regular, na qual os \'{a}tomos  executam pequenas vibra\c{c}\~{o}es em torno de posi\c{c}\~{o}es relativas mais ou menos fixas.} como um g\'{a}s de b\'{o}sons n\~{a}o massivos, chamados \textit{f\^{o}nons}, baseia-se na aproxima\c{c}\~{a}o harm\^{o}nica da energia potencial efetiva,  a qual descreve as intera\c{c}\~{o}es de cada  \'{a}tomo  com seus vizinhos. Desse modo, cada \'{a}tomo \'{e} associado a um oscilador harm\^{o}nico independente dos demais, cujo espectro de energia, $\epsilon_n(\nu$), segundo a Mec\^{a}nica Qu\^{a}ntica, \'{e} dado por
 $$ \epsilon_n (\nu) = ( n + 1/2) \, h \nu  = \epsilon_\circ + n h \nu \qquad (n=0,1, \ldots ),$$
em que $\nu$ \'{e} a frequ\^{e}ncia natural de vibra\c{c}\~{a}o, e $\epsilon_\circ = h \nu/2$ \'{e} a energia do estado fundamental de um particular oscilador.

A energia de cada \'{a}tomo em um s\'{o}lido cristalino em equil\'{\i}brio t\'{e}rmico, portanto,  \'{e} composta por uma parcela constante, a energia do estado fundamental,\footnote{~Como a energia \'{e} definida a menos de uma constante, essa parcela pode ser considera nula.} e uma parcela que depende do grau de excita\c{c}\~{a}o do \'{a}tomo, ou seja, da temperatura do cristal.

Considerando que cada estado excitado de um oscilador corresponde a $n$ part\'{\i}culas independentes n\~{a}o massivas, cada qual com energia $ \epsilon = h \nu$,
% e {\it momentum} $ p = \epsilon/c$,
%\footnote{~De acordo com a rela\c{c}\~{a}o de L.~de Broglie  \cite{CARUSO-OGURI}, o comprimento de onda, a frequ\^{e}ncia, a velocidade e o {\it momentum} de uma part\'{\i}cula est\~{a}o relacionados por $$ \lambda = h/p=c/\nu$$ e, portanto,  esses osciladores n\~{a}o possuem massa. }
o conjunto de osciladores que representam as vibra\c{c}\~{o}es at\^{o}micas  do cristal pode ser associado a um sistema de part\'{\i}culas independentes que se comportam como um g\'{a}s de b\'{o}sons n\~{a}o massivos, pois o n\'{u}mero ($n$) de part\'{\i}culas associado a cada estado depende da temperatura e, portanto, n\~{a}o obedece ao princ\'{\i}pio de exclus\~{a}o de Pauli.

\begin{center}
\fbox{
\begin{minipage}{13cm}
\baselineskip=10pt

\vspace*{.1cm}
  Em baixas temperaturas, as vibra\c{c}\~{o}es at\^{o}micas em um s\'{o}lido cristalino s\~{a}o equivalentes a um g\'{a}s degenerado de b\'{o}sons n\~{a}o massivos, os f\^{o}nons.

\vspace*{.2cm}
\end{minipage}
}
\end{center}

Como cada \'{a}tomo corresponde a 3 osciladores independentes, o n\'{u}mero total de osciladores no cristal \'{e} $3N$, e
 %o n\'{u}mero e a  densidade de estados para os f\^{o}nons s\~{a}o  dados por
%$$ G(\epsilon) =  3 N\, \left(\frac{\displaystyle \epsilon}{\displaystyle kT_{_D} } \right)^3  \qquad \mbox{e} \qquad  g(\epsilon) = \frac{\displaystyle 9 N}{\displaystyle (kT_{_D})^3} \, \epsilon^2 .$$
 o calor espec\'{\i}fico molar \'{e} dado pela lei de Debye \cite{CARUSO-OGURI, FERMI1, SOMMER, ZEMANSKY, MARIO, Black},\footnote{~A rigor a lei de Debye \'{e} v\'{a}lida em temperaturas menores que $T_{_D}/50$ \cite{Black}.}
\begin{equation} \label{Debye_lei}
\fbox{~$ \displaystyle
 \displaystyle
c_{_V} =  12 R \alpha \left( \frac{T}{T_{_D}} \right)^3  \quad (T \ll T_{_D})  $~} \qquad  \mbox{\small (f\^{o}nons).}
 \end{equation}
%Apesar de o n\'{u}mero de f\^{o}nons em um cristal n\~{a}o ser definido, pois depende da temperatura,   a energia do estado fundamental \`{a} $0$~K  tem valor constante e finito  determinado pelo n\'{u}mero de \'{a}tomos do cristal.

\begin{figure}[hbtp]
\centerline{\includegraphics[width=9.5cm]{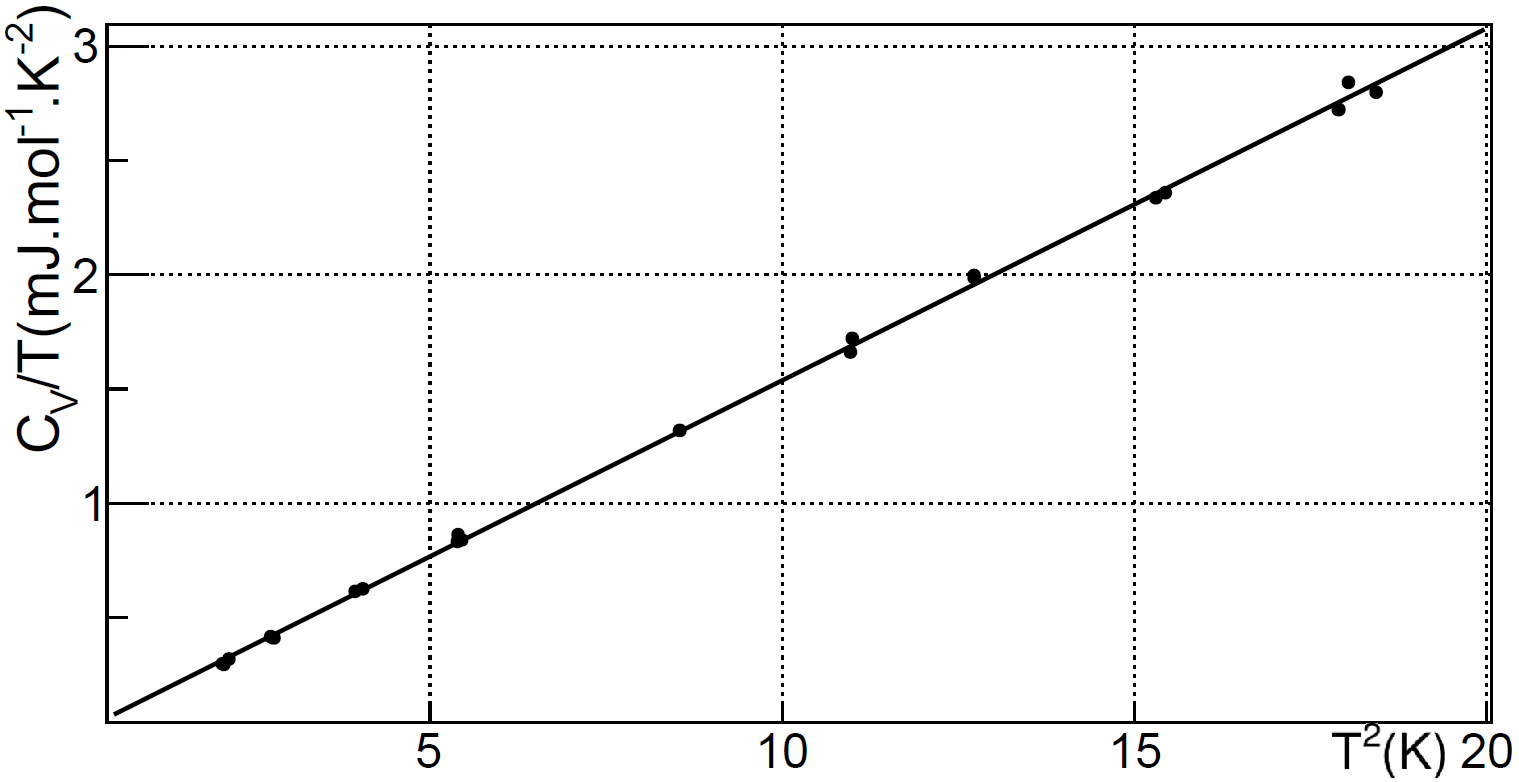}}
\caption{Calor espec\'{\i}fico molar do  {\tt KCl}  em temperaturas abaixo de 4~K. A linha cont\'{\i}nua \'{e} a reta de ajuste aos dados (pontos)  de Keesom e Pearlman, de coeficiente angular $15,\!4 \times 10^{-5}$~J$\cdot$mol$^{-1}\cdot$K$^{-2}$. A temperatura de Debye estimada \'{e} da ordem de $(233 \pm 3)$~K.}
\label{c_solido}
\end{figure}

   O resultado do experimento pioneiro de P.~H.~Keesom e N.~Pearlman \cite{KEESOM}, de 1955, apresentado no gr\'{a}fico $c_{_V}/T$  {\it versus}  $T^2$  da figura~\ref{c_solido}, mostra que  o comportamento do calor espec\'{\i}fico dos s\'{o}lidos n\~{a}o met\'{a}licos  a baixas temperaturas  ($T < T_{_D}/50 \sim 4$~K) \'{e}  compat\'{\i}vel com a lei de Debye.

Em altas temperaturas  ($T \gg T_{_D}$), os s\'{o}lidos comportam-se como um sistema n\~{a}o degenerado cujo calor espec\'{\i}fico molar ($c_{_V}$) obedece \`{a} lei de Dulong-Petit \cite{DULONG, CARUSO-OGURI, FERMI1, SOMMER, ZEMANSKY, MARIO, Black},
\begin{equation}\label{calor5}
\fbox{~$ \displaystyle
c_{_V} = 3R \simeq 6~{\rm cal}\cdot{\rm mol}^{-1}\cdot{\rm K}^{-1} \simeq 25~{\rm J}\cdot{\rm mol}^{-1}\cdot{\rm K}^{-1}  \qquad  (T \gg T_{_D})  $~} \qquad  \mbox{\small (f\^{o}nons).}
 \end{equation}

\begin{figure}[hbtp]
\centerline{\includegraphics[width=12cm]{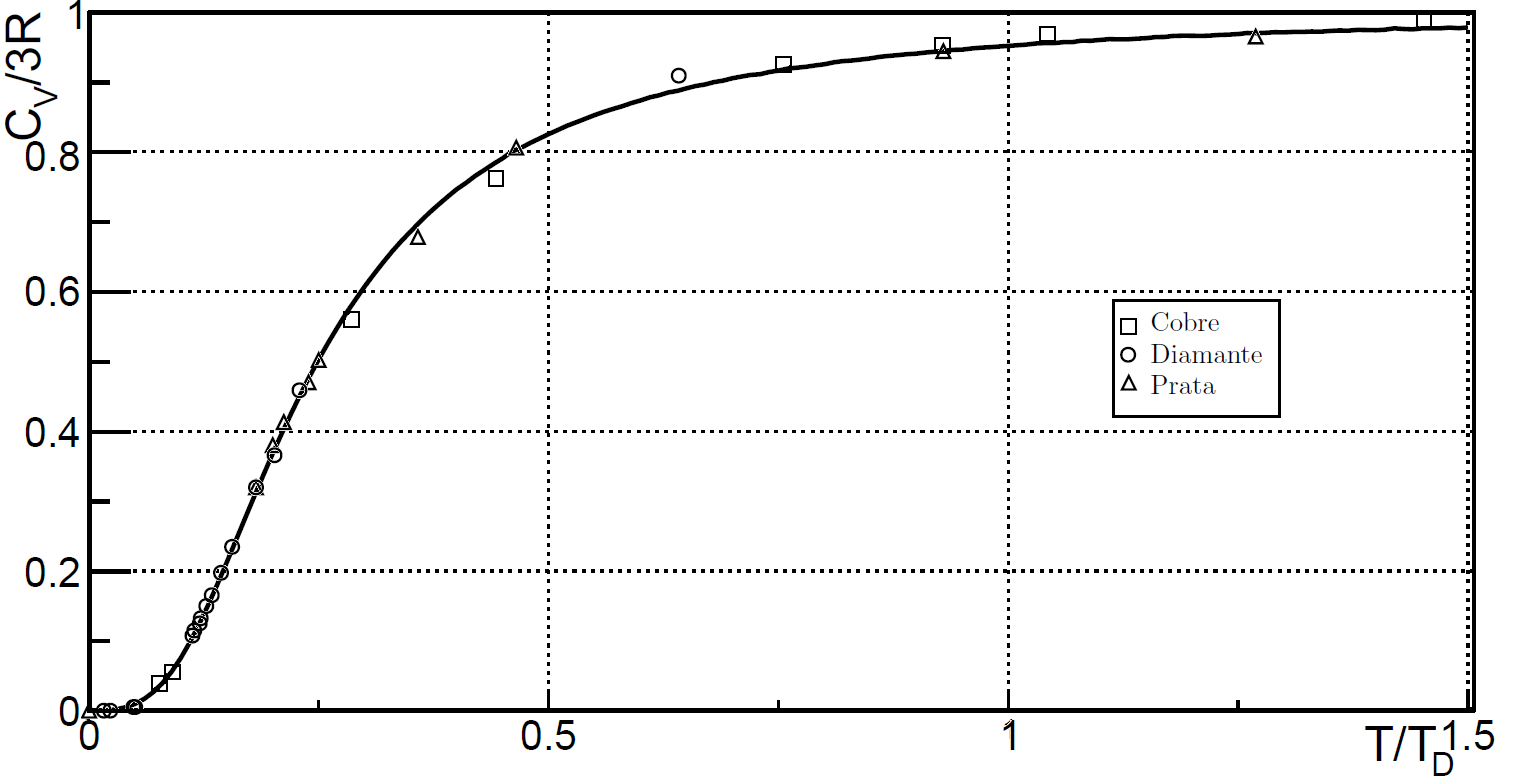}}
\caption{Compara\c{c}\~{a}o da curva te\'{o}rica de Debye (linha cont\'{\i}nua) com os dados experimentais  (pontos) relativos ao calor espec\'{\i}fico molar de alguns s\'{o}lidos \cite{MARIO}. }
\label{debye}
\end{figure}

 Tanto do ponto de vista  te\'{o}rico, como experimentalmente, o calor espec\'{\i}fico dos s\'{o}lidos cresce suavemente com a temperatura at\'{e} o valor limite dado pela lei de Dulong-Petit (figura~\ref{debye}).

%Originalmente, o m\'{e}todo de Debye \cite{CARUSO-OGURI, SOMMER,  MARIO, Black} tem por hip\'{o}tese que em baixas frequ\^{e}ncias as oscila\c{c}\~{o}es at\^{o}micas est\~{a}o  associadas \`{a} propaga\c{c}\~{a}o de ondas ac\'{u}sticas em um meio s\'{o}lido el\'{a}stico e isotr\'{o}pico. Segundo M.~Blackman~\cite{Black}, devido ao seu enorme \^{e}xito inicial, na compara\c{c}\~{a}o com diversos dados experimentais, tornou-se   um exemplo daquilo que se pode chamar {\it canoniza\c{c}\~{a}o a priori} de uma teoria. No entanto, quando  aplicada ao \'{u}nico cristal c\'{u}bico (tungst\^{e}nio) para o qual a condi\c{c}\~{a}o de isotropia \'{e} satisfeita, n\~{a}o se ajusta aos dados.

%Por  outro lado,  \`{a} mesma \'{e}poca  de Debye, Born e von K\'{a}rm\'{a}n (1912), em vez de considerarem o s\'{o}lido como um meio el\'{a}stico cont\'{\i}nuo,  atacaram esse problema de modo din\^{a}mico,  obtendo algumas rela\c{c}\~{o}es de dispers\~{o}es apropriadas para descrever as vibra\c{c}\~{o}es em uma rede cristalina.

\subsubsection{O calor espec\'{\i}fico dos metais }

%\paragraph*{}
 Os metais constituem uma classe especial de s\'{o}lidos cristalinos. Al\'{e}m dos \'{\i}ons que constituem a rede cristalina, e vibram em torno de suas posi\c{c}\~{o}es de equil\'{\i}brio, possuem tamb\'{e}m praticamente o mesmo n\'{u}mero de el\'{e}trons, os el\'{e}trons de condu\c{c}\~{a}o, que n\~{a}o est\~{a}o associados a nenhum particular \'{\i}on.

% , que se deslocam pelo cristal como um g\'{a}s degenerado de f\'{e}rmions de {\it spin 1/2}.

A hip\'{o}tese de que algumas das propriedades de um metal pudessem ser obtidas a partir do modelo do g\'{a}s de el\'{e}trons remontam \`{a} \'{e}poca de P.~Drude (1900). Tal hip\'{o}tese apoia-se no fato de que em um cristal os \'{\i}ons positivos do metal estabelecem um campo eletromagn\'{e}tico que tende a anular a a\c{c}\~{a}o dos outros el\'{e}trons sobre um determinado el\'{e}tron. No entanto, ao se considerar os el\'{e}trons de condu\c{c}\~{a}o como um  g\'{a}s  n\~{a}o degenerado, os resultados obtidos n\~{a}o foram compat\'{\i}veis com o comportamento t\'{e}rmico observado. Apenas quando Sommerfeld  considerou-os como um g\'{a}s degenerado de f\'{e}rmions de {\it spin} 1/2, os resultados te\'{o}ricos tornaram-se compat\'{\i}veis com os experimentos~\cite{CORAK}.

 Assim,  o calor espec\'{\i}fico  de um metal possui uma componente devida \`{a}s vibra\c{c}\~{o}es da rede, ou a  um g\'{a}s degenerado de f\^{o}nons, e  outra devida ao g\'{a}s degenerado de el\'{e}trons de condu\c{c}\~{a}o do metal,
 $$ c_{\mbox{\tiny metal}} \ = \  c_{\mbox{\tiny f\^{o}nons}} \ + \ c_{\mbox{\tiny el\'{e}trons}}. $$

Como  \`{a} temperatura ambiente  $T \ll T_{_F}$  para qualquer  metal, a  contribui\c{c}\~{a}o eletr\^{o}nica  s\'{o} \'{e} relevante em temperaturas muito menores que a temperatura  de Debye do metal, quando o calor espec\'{\i}fico pode ser expresso como
  $$ c_{\mbox{\tiny metal}} \ = \ \alpha \ T^3 \ + \ \gamma \ T   \qquad  ( T \ll T_{_D}), $$
\noindent
em que o par\^{a}metros $\alpha$ e $\gamma$ determinam, respectivamente,  a temperatura de Debye  e a  temperatura de Fermi.

\begin{figure}[hbtp]
\centerline{\includegraphics[width=10cm]{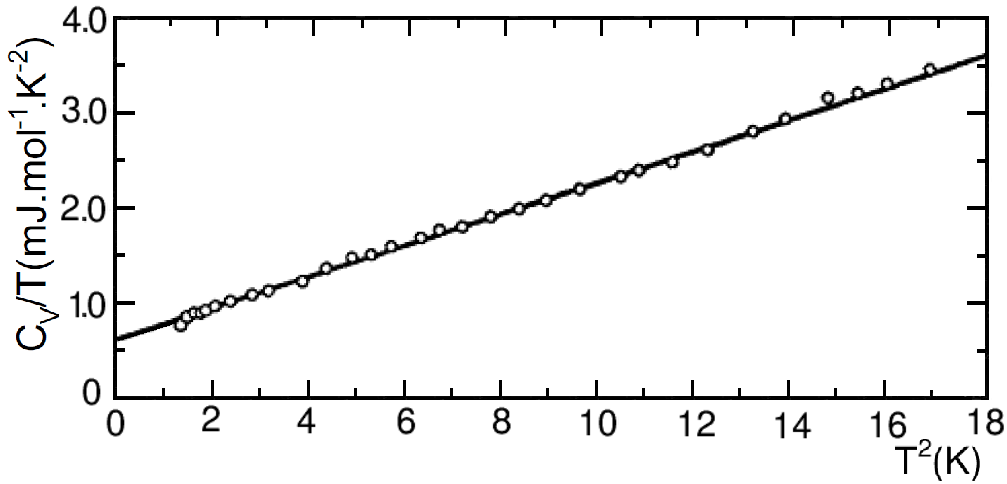}}
\caption{Calor espec\'{\i}fico de uma amostra de  prata. A linha cont\'{\i}nua  \'{e} a reta de ajuste aos dados (pontos) experimentais de Corak {\it et. al}. O coeficiente linear da ordem de $0,\!6 \times 10^{-3}$~J$\cdot$mol$^{-1}\cdot$K$^{-2}$ corresponde  \`{a} temperatura de Fermi de $6,\!8 \times 10^{4}$~K.}
\label{metal}
\end{figure}

Esse comportamento dos metais, j\'{a} observado por Keesom (1935), foi estabelecido nos experimentos realizados por W.~Corak, M.~P.~Garfunkel, C.~B.~Satterthwaite e  A.~Wexler em 1955 \cite{CORAK}  (figura~\ref{metal}).

\subsubsection{A radia\c{c}\~{a}o de corpo negro} \label{Planck}

%\paragraph*{}
De maneira an\'{a}loga \`{a}s vibra\c{c}\~{o}es, o campo eletromagn\'{e}tico associado \`{a} radia\c{c}\~{a}o de corpo negro em equil\'{\i}brio t\'{e}rmico pode ser  representado por um g\'{a}s de b\'{o}sons n\~{a}o massivos -- os  \textit{f\'{o}tons} com energias $h\nu  \  (\nu=0,\ldots\infty)$ \cite{CARUSO-OGURI, FERMI1, SOMMER, MARIO, STUDART}.

Nas  vibra\c{c}\~{o}es at\^{o}micas, a energia do estado fundamental a $0$~K  tem valor constante e finito determinado pelo n\'{u}mero de osciladores do cristal; no caso da radia\c{c}\~{a}o de corpo negro, o n\'{u}mero total de osciladores associados ao campo eletromagn\'{e}tico n\~{a}o \'{e} finito. Assim, para se contornar o problema de um n\'{u}mero infinito de termos contribuir para a energia do estado fundamental  do sistema a 0~K, considera-se que a energia do sistema \'{e} a energia dos estados excitados, ou seja, a energia dos f\'{o}tons.
%Esses estados est\~{a}o associados a um n\'{u}mero  n\~{a}o definido  de f\'{o}tons.

Uma vez que n\~{a}o existem restri\c{c}\~{o}es quanto ao n\'{u}mero de f\'{o}tons, pois  esse  n\'{u}mero  pode ser infinito, a radia\c{c}\~{a}o de corpo negro sempre constitui um  sistema degenerado de b\'{o}sons n\~{a}o massivos em qualquer temperatura.
%Ou seja, o g\'{a}s de f\'{o}tons sempre est\'{a} abaixo de sua temperatura cr\'{\i}tica.

Substituindo-se a express\~{a}o para a temperatura de Debye (a segunda das eqs.~(\ref{Tc1})  na eq.~(\ref{U3}),
 a densidade de energia ($u = U/V$) da radia\c{c}\~{a}o de corpo negro pode ser escrita na forma usual da lei de Stefan-Boltzmann \cite{CARUSO-OGURI, FERMI1, SOMMER, MARIO},
\begin{equation}\label{SB}
  u = \frac{8 \pi \alpha}{3} \, \frac{k^4}{(ch)^3} \, T^4 = a \, T^4,
 \end{equation}

\noindent
sendo $a \simeq  7,\!6 \times 10^{-16}~{\rm J}\cdot{\rm m}^{-3}\cdot{\rm K}^{-4}$, para $\alpha= \pi^4/5$.

%O problema da radia\c{c}\~{a}o do corpo negro, al\'{e}m de ter sido um dos principais germes para a cria\c{c}\~{a}o e constru\c{c}\~{a}o  da Mec\^{a}nica Qu\^{a}ntica Moderna, foi tamb\'{e}m  o principal modelo sobre o qual se deu a afirma\c{c}\~{a}o da pr\'{o}pria Mec\^{a}nica Estat\'{\i}stica.

%O desenvolvimento paralelo dessas teorias, no in\'{\i}cio de s\'{e}culo XX, deve-se, n\~{a}o s\'{o} aos aspectos probabil\'{\i}sticos de ambas as teorias mas, tamb\'{e}m ao fato de que o comportamento dos sistemas estudados devem estar sujeito, em \'{u}ltima an\'{a}lise, \`{a}s leis da Mec\^{a}nica Qu\^{a}ntica.

\subsubsection{B\'{o}sons massivos n\~{a}o relativ\'{\i}sticos fortemente degenerados e  condensa\c{c}\~{a}o de Bose-Einstein}

%\paragraph*{}
%A diferen\c{c}a entre o comportamento dos b\'{o}sons n\~{a}o massivos e massivos, com rela\c{c}\~{a}o ao n\'{u}mero de part\'{\i}culas excitadas, decorre principalmente do fen\^{o}meno de cria\c{c}\~{a}o e aniquila\c{c}\~{a}o de
O n\'{u}mero de part\'{\i}culas excitadas em um g\'{a}s de b\'{o}sons \'{e} diferente quando eles s\~{a}o massivos ou n\~{a}o, devido, principalmente, ao fen\^{o}meno de cria\c{c}\~{a}o e aniquila\c{c}\~{a}o de part\'{\i}culas. Tanto a energia da  radia\c{c}\~{a}o eletromagn\'{e}tica de corpo negro, quanto das oscila\c{c}\~{o}es at\^{o}micas nos cristais, podem ser descritas pela soma das energias de  b\'{o}sons n\~{a}o massivos, respectivamente   denominados f\'{o}tons e f\^{o}nons.  Uma vez que a quantidade desses  b\'{o}sons n\~{a}o massivos n\~{a}o \'{e} fixa, dependendo fortemente da temperatura ($\propto T^3$),   diz-se que f\'{o}tons e f\^{o}nons podem ser criados ou aniquilados.
 %o n\'{u}mero de b\'{o}sons n\~{a}o massivos criados com energias acima da energia do estado fundamental em  temperaturas acima de 0~K \'{e} muito maior do que aqueles  que s\~{a}o excitados  no caso de gases de b\'{o}sons massivos.
%\footnote{~No caso de b\'{o}sons n\~{a}o relativ\'{\i}sticos n\~{a}o h\'{a} cria\c{c}\~{a}o ou aniquila\c{c}\~{a}o de part\'{\i}culas.}

Diferentemente dos  b\'{o}sons n\~{a}o massivos, os b\'{o}sons massivos n\~{a}o relativ\'{\i}sticos n\~{a}o podem ser criados ou aniquilados, o que implica conserva\c{c}\~{a}o do n\'{u}mero de part\'{\i}culas.\footnote{~No caso de b\'{o}sons massivos relativ\'{\i}sticos, deve-se considerar tamb\'{e}m os processos de cria\c{c}\~{a}o e aniquila\c{c}\~{a}o.} Desse modo, a energia do estado fundamental \'{e} nula e pode-se admitir que em baixas temperaturas o n\'{u}mero de part\'{\i}culas excitadas com energia m\'{e}dia da ordem de $kT$ seja dado pelo n\'{u}mero de estados com energia at\'{e} $kT$,\footnote{~A temperatura cr\'{\i}tica e a densidade de estados para b\'{o}sons massivos n\~{a}o relativ\'{\i}sticos s\~{a}o calculadas do mesmo modo que para os f\'{e}rmions n\~{a}o relativ\'{\i}sticos, pois a rela\c{c}\~{a}o entre a energia ($\epsilon$) e o {\it momentum} ($p$) de part\'{\i}culas n\~{a}o relativ\'{\i}sticas de massa $m$ \'{e} dada por $p = \sqrt{2m\epsilon}$.}
\begin{equation}\label{N0}
N_{\epsilon > 0} = \int_0^{kT}
\frac{3}{2} \frac{N}{(kT_c)^{3/2}}  \epsilon^{1/2} \, \mbox{d}\epsilon =
N \left( \frac{T}{T_c} \right)^{3/2},
\end{equation}
\noindent
e a energia interna e o calor espec\'{\i}fico molar a volume constante, compat\'{\i}vel com a 3\na\ lei da Termodin\^{a}mica, sejam, aproximadamente,
$$
 \displaystyle
U \simeq N_{\epsilon > 0} \,  kT =  N k \, \frac{T^{5/2}}{T_c^{3/2}},
\qquad \mbox{e} \qquad
c_{_V} \simeq     R \, \left( \frac{T}{T_c} \right)^{3/2}.
$$

O  n\'{u}mero de b\'{o}sons  no estado fundamental  ($N_\circ = N - N_{\epsilon > 0}$), por sua vez, \'{e} dado por
\begin{equation}\label{NB}
    N_\circ = N \left[ 1 - \left( \frac{T}{T_c} \right)^{3/2} \right].
\end{equation}

\begin{figure}[hbtp]
\centerline{\includegraphics[width=4.5cm]{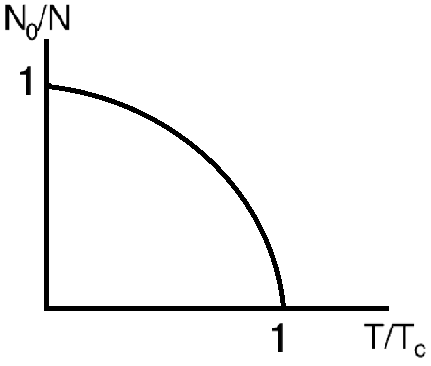}}
\caption{N\'{u}mero de b\'{o}sons n\~{a}o massivos no condensado de Bose-Einstein.}
\label{c_einstein}
\end{figure}

 De acordo com a eq.~(\ref{NB}), abaixo da temperatura cr\'{\i}tica o g\'{a}s se aproxima de um estado em que o n\'{u}mero de b\'{o}sons massivos agrupados nas vizinhan\c{c}as do estado fundamental cresce rapidamente (figura~\ref{c_einstein}). Com base nesse comportamento, Einstein sup\~{o}e que na temperatura cr\'{\i}tica ocorra um novo fen\^{o}meno, no qual o sistema de b\'{o}sons massivos atinja um estado mais organizado da mat\'{e}ria, em uma transi\c{c}\~{a}o do tipo desordem--ordem. Em analogia com a condensa\c{c}\~{a}o de um g\'{a}s ordin\'{a}rio, o fen\^{o}meno \'{e} conhecido, desde ent\~{a}o, como condensa\c{c}\~{a}o de Bose-Einstein \cite{LONDON, DAHMEN2, BAGNATO1}.

Sendo nula a energia do estado fundamental, a grande maioria das part\'{\i}culas do chamado condensado de Bose-Einstein possui {\it momentum} nulo em baixas temperaturas ($T/T_c < 0,\!4$). Em maior n\'{u}mero que as demais, tais  part\'{\i}culas contribuiriam com maior peso para as propriedades macrosc\'{o}picas do sistema, como press\~{a}o e viscosidade.

Essas conclus\~{o}es  foram estabelecidas por Einstein em 1924  e 1925 \cite{EINSTEIN1, EINSTEIN2}, a partir do m\'{e}todo  de contagem de f\'{o}tons da radia\c{c}\~{a}o de corpo negro utilizado por Bose.  Bose considerou que os f\'{o}tons n\~{a}o obedeciam ao princ\'{\i}pio de Pauli, \`{a} \'{e}poca desconhecido, e eram indistinguiveis.\footnote{~ Bose tamb\'{e}m admitiu corretamente que, devido a polariza\c{c}\~{a}o da luz,  o fator  de multiplica\c{c}\~{a}o de estados era igual a 2  ($f=2$).} Ao estender o procedimento de Bose aos gases  moleculares, Einstein obt\'{e}m uma teoria qu\^{a}ntica dos gases ideais degenerados de b\'{o}sons massivos n\~{a}o relativ\'{\i}sticos.

A descoberta da exist\^{e}ncia de um novo estado condensado da mat\'{e}ria foi considerada, inicialmente, apenas um resultado matem\'{a}tico, sem possibilidade de verifica\c{c}\~{a}o. Devido ao baix\'{\i}ssimo valor da temperatura cr\'{\i}tica de um g\'{a}s molecular ($T_c < 0,\!1$~K), qualquer g\'{a}s real a t\~{a}o baixa temperatura estaria no estado l\'{\i}quido.

Apesar dos argumentos contr\'{a}rios \`{a} condensa\c{c}\~{a}o de Bose-Einstein, F.~London, em 1938, estabelece que o calor espec\'{\i}fico molar a volume constante de um g\'{a}s de b\'{o}sons massivos n\~{a}o relativ\'{\i}sticos para temperaturas at\'{e} o valor  cr\'{\i}tico \'{e} dado por~\cite{LONDON}
\begin{equation}\label{London_c}
\fbox{~$ \displaystyle
c_{_V} = 1,\!93 \, R \, \left( \frac{T}{T_c} \right)^{3/2} \quad (T \leq T_c)$~} \quad  \mbox{\small ( b\'{o}sons massivos n\~{a}o relativ\'{\i}sticos).}
\end{equation}

\begin{figure}[hbtp]
\centerline{\includegraphics[width=9.5cm]{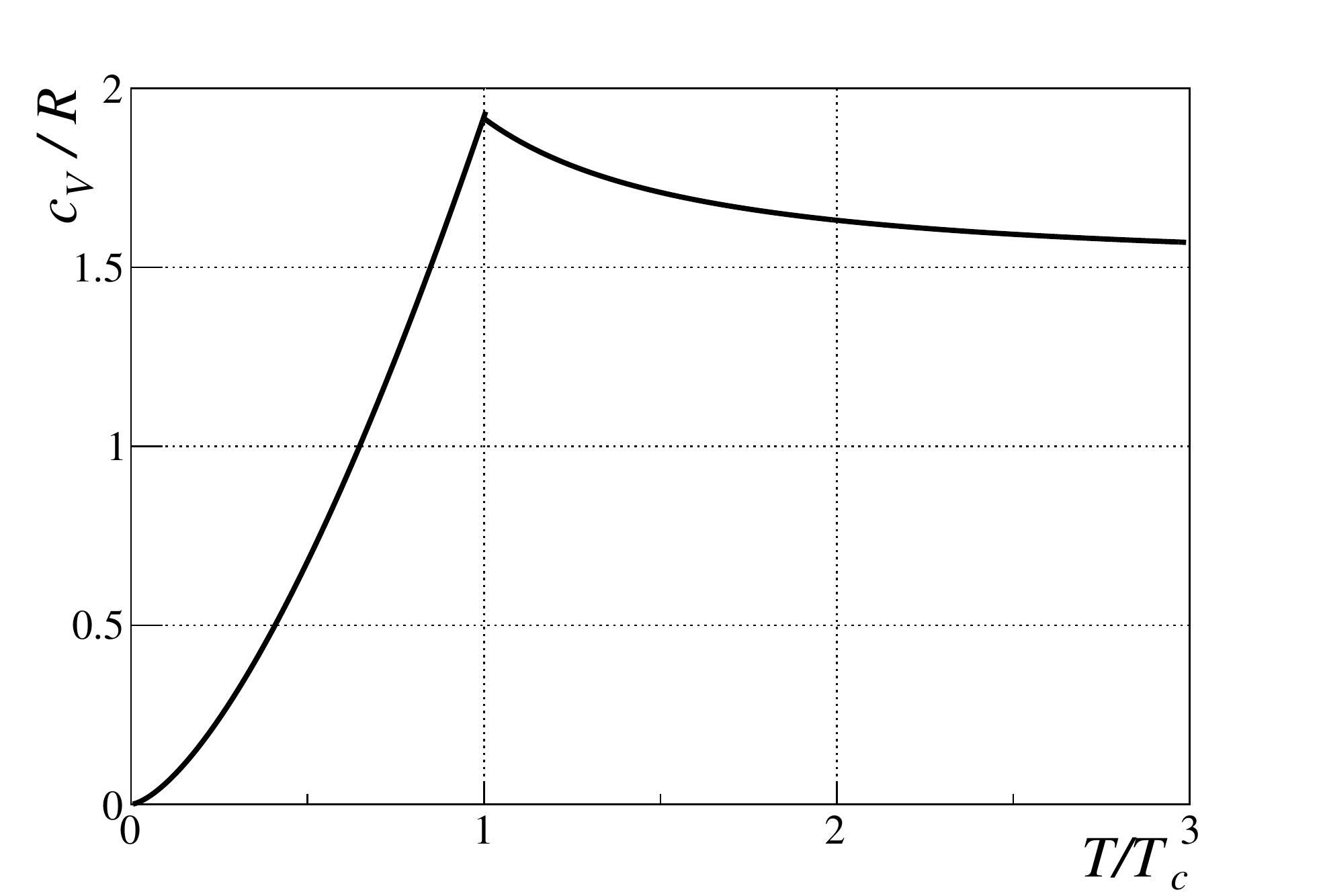}}
\caption{Calor espec\'{\i}fico de um g\'{a}s de b\'{o}sons massivos n\~{a}o relativ\'{\i}sticos.}
\label{c_london}
\end{figure}

Como o valor do calor espec\'{\i}fico molar \`{a} temperatura cr\'{\i}tica  ($1,\!93\, R$) excede o valor cl\'{a}ssico ($1,\!5\, R$) para o qual deve se aproximar assintoticamente para $T > T_c$, nas vizinhan\c{c}as da temperatura cr\'{\i}tica o calor espec\'{\i}fico do g\'{a}s apresenta um comportamento n\~{a}o suave (figura~\ref{c_london}), o que implica descontinuidade em sua derivada.

%Segundo L.~Landau, uma transi\c{c}\~{a}o de fase \'{e} sempre acompanhada da mudan\c{c}a brusca de par\^{a}metros que  caracterizam a  ordem de um sistema.

Foi exatamente a partir desse comportamento que London  defendeu a hip\'{o}tese de Einstein sobre a temperatura cr\'{\i}tica de um g\'{a}s de b\'{o}sons massivos n\~{a}o relativ\'{\i}sticos estar associada a uma transi\c{c}\~{a}o de fase. Rebatendo as cr\'{\i}ticas de que o fen\^{o}meno da condensa\c{c}\~{a}o de Bose-Einstein resultaria de um artif\'{\i}cio matem\'{a}tico, London sugere a transi\c{c}\~{a}o l\'{\i}quido-superfluido do h\'{e}lio ({\tt He}$^4$) como um exemplo de condensa\c{c}\~{a}o de Bose-Einstein que ocorreria na natureza.\footnote{~Apesar de ocorrer tamb\'{e}m a concentra\c{c}\~{a}o de part\'{\i}culas no estado fundamental de um g\'{a}s de b\'{o}sons n\~{a}o massivos em baixas temperaturas, a energia do estado fundamental n\~{a}o \'{e} nula e o calor espec\'{\i}fico varia suavemente (figura~\ref{debye}), n\~{a}o havendo descontinuidade em sua derivada  e, portanto, n\~{a}o caracterizando uma transi\c{c}\~{a}o de fase.}

\subsubsection{O  {\tt \bf  He}$^4$ l\'{\i}quido}

%\paragraph*{}
A particularidade mais marcante do   {\tt   He}$^4$  l\'{\i}quido ($T \sim 4$~K)  quando \'{e} resfriado em equil\'{\i}brio com  sua press\~{a}o de vapor, ap\'{o}s alcan\c{c}ar a temperatura cr\'{\i}tica da ordem de 2~K, \'{e} a s\'{u}bita capacidade de fluir em tubos capilares sem exercer press\~{a}o, como se deixasse de ter  viscosidade. Essa mudan\c{c}a brusca na viscosidade, acompanhada tamb\'{e}m de uma varia\c{c}\~{a}o brusca no calor espec\'{\i}fico, \'{e} a propriedade que  caracteriza o fen\^{o}meno como uma transi\c{c}\~{a}o de fase l\'{\i}quido-superfluido.

Uma vez que os \'{a}tomos de {\tt He}$^4$  possuem seis part\'{\i}culas (2 pr\'{o}tons, 2 n\^{e}utrons e 2 el\'{e}trons) de {\it spin} semi-inteiros, F.~London sup\~{o}e, em primeira aproxima\c{c}\~{a}o, que o h\'{e}lio l\'{\i}quido constitu\'{\i}do dos is\'{o}topos {\tt He}$^4$  pode ser considerado um sistema de  part\'{\i}culas independentes de {\it spin} inteiros, ou seja, como um g\'{a}s de b\'{o}sons massivos  n\~{a}o relativ\'{\i}sticos, e que a transi\c{c}\~{a}o  l\'{\i}quido-superfluido tivesse rela\c{c}\~{a}o com o fen\^{o}meno de condensa\c{c}\~{a}o de Bose-Einstein.

Essa hip\'{o}tese \'{e} refor\c{c}ada quando, ao considerar-se {\tt  He}$^4$ l\'{\i}quido como um g\'{a}s de b\'{o}sons, o valor da temperatura cr\'{\i}tica \'{e} cerca de $3$~K, conforme Tabela~\ref{Tc}. Apesar da similaridade e do estabelecimento posterior do fen\^{o}meno como um exemplo de condensa\c{c}\~{a}o de Bose-Einstein,\footnote{~A observa\c{c}\~{a}o da popula\c{c}\~{a}o de \'{a}tomos de {\tt  He}$^4$  com {\it momentum} nulo foi realizada  em 1982 \cite{BAGNATO1}.} por n\~{a}o ser exatamente um sistema de part\'{\i}culas que n\~{a}o interagem, o comportamento do calor espec\'{\i}fico do  {\tt   He}$^4$ l\'{\i}quido \'{e} muito diferente do g\'{a}s ideal degenerado de b\'{o}sons massivos.

A procura de um sistema gasoso que exibisse a condensa\c{c}\~{a}o de Bose-Einstein s\'{o} teve \^{e}xito em 1995, quando vapores de  \'{a}tomos de {\tt Rb}, a uma densidade de $10^{12}$~cm$^{-3}$, foram resfriados a baix\'{\i}ssima temperatura, cerca de 100~nK, por um grupo do JILA.\footnote{~Joint Institute for Laboratoty Astrophysics, um instituto conjunto da University of Colorado e do NIST (National Institute of Standards and Technology).} Desde ent\~{a}o, v\'{a}rios  experimentos sobre esse estranho comportamento da mat\'{e}ria foram realizados, constituindo-se em uma ativa \'{a}rea de pesquisa da F\'{\i}sica de ultra baixas temperaturas \cite{BAGNATO1, BAGNATO2}.

\section{Considera\c{c}\~{o}es finais}

%\paragraph*{}
Do ponto de vista din\^{a}mico, ainda que em equil\'{\i}brio t\'{e}rmico, o n\'{u}mero de part\'{\i}culas associado a cada n\'{\i}vel de energia ($\epsilon$) de um g\'{a}s varia incessantemente. Devido a essas flutua\c{c}\~{o}es e ao grande n\'{u}mero de part\'{\i}culas, a distribui\c{c}\~{a}o das part\'{\i}culas pelos estados associados aos n\'{\i}veis de energia do g\'{a}s s\~{a}o caracterizadas pelos n\'{u}meros m\'{e}dios de ocupa\c{c}\~{a}o, $ \langle n _\epsilon \rangle$, ou popula\c{c}\~{a}o m\'{e}dia dos estados.

A distribui\c{c}\~{a}o das part\'{\i}culas depende do estado de degeneresc\^{e}ncia do g\'{a}s e, portanto, da natureza de suas part\'{\i}culas constituintes. Al\'{e}m das correla\c{c}\~{o}es decorrentes do princ\'{\i}pio de exclus\~{a}o de Pauli, deve-se considerar que as part\'{\i}culas, b\'{o}sons e f\'{e}rmions, de um g\'{a}s ideal s\~{a}o, entre si, indistingu\'{\i}veis.

As tr\^{e}s distribui\c{c}\~{o}es usuais para os gases ideais podem ser sintetizadas como \cite{FERMI1,SOMMER}
\begin{equation} \label{pop}
\fbox{
$~~ \displaystyle
  \langle n_\epsilon  \rangle \ =\
 \frac{1}{\lambda^{ \scriptsize - 1} e^{\epsilon/kT} +a}\, ,~~$}
\end{equation}
em que $\lambda$ \'{e} a fugacidade do g\'{a}s.\footnote{~Para part\'{\i}culas massivas a fugacidade ($\lambda$) \'{e} determinada pela restri\c{c}\~{a}o ao n\'{u}mero de part\'{\i}culas; para b\'{o}sons n\~{a}o massivos n\~{a}o h\'{a} restri\c{c}\~{a}o associada ao n\'{u}mero de part\'{\i}culas, e  no caso dos f\^{o}nons, o espectro de energia tem um limite determinado pela temperatura de Debye ($\epsilon_{\mbox{\tiny max}} = k T_{_D}$).}

A {distribui\c{c}\~{a}o de Maxwell-Boltzmann} ($a=0$) descreve o comportamento da popula\c{c}\~{a}o m\'{e}dia em um g\'{a}s ideal n\~{a}o degenerado, a {distribui\c{c}\~{a}o de Fermi-Dirac} ($a=1$), a popula\c{c}\~{a}o m\'{e}dia em um g\'{a}s ideal de f\'{e}rmions, e a {distribui\c{c}\~{a}o de Bose-Einstein} ($a=-1$), a popula\c{c}\~{a}o m\'{e}dia em gases ideais de b\'{o}sons n\~{a}o massivos ($\lambda=1$) e massivos.\footnote{~Para b\'{o}sons n\~{a}o massivos, principalmente, para f\'{o}tons da radia\c{c}\~{a}o de corpo negro, a distribui\c{c}\~{a}o \'{e} chamada tamb\'{e}m distribui\c{c}\~{a}o de Planck.}

Assim, as propriedades macrosc\'{o}picas de um g\'{a}s em equil\'{\i}brio t\'{e}rmico, como a energia m\'{e}dia $U$, ou energia interna, e o n\'{u}mero total de part\'{\i}culas $N$, satisfazem as seguintes rela\c{c}\~{o}es
\begin{equation} \label{UN}
 \left\{
\begin{array}{l}
U =  \displaystyle \int_0^{\infty} \epsilon \, g(\epsilon) \,
 \langle n_\epsilon \rangle  \, \mbox{d}\epsilon,     \\
\vspace*{.2cm} \\
N = \displaystyle \int_0^{\infty} g(\epsilon) \, \langle n_\epsilon \rangle \, \mbox{d} \epsilon,
 \end{array}
\right.
\end{equation}
sendo  $g(\epsilon)$  a  densidade de estados.

 A partir dessas express\~{o}es, eqs.~(\ref{UN}), s\~{a}o obtidas estimativas mais acuradas sobre o comportamento dos gases, correspondentes a um amplo dom\'{\i}nio de temperaturas, desde as mais baixas ($T \ll T_c$) \`{a}s mais altas ($T \gg T_c$).

%\def\ra{-\kern-.4em\raise.8ex\hbox{{\tt \tiny a}}\ }

%\na

% \raise-.8ex\hbox{$\tilde{\hspace{6pt}}$}dia
%\def\til{~\kern-.4em\raise.8ex\hbox{a}\ }
%\def\til{~\kern-.1em\raise-.8ex\hbox{$\tilde{\hspace{0pt}}$}\ }
\def\til{~\kern-.4em\raise-.8ex\hbox{\~{\hspace{-4pt}}}\ }

%d\til{a}ge

%\newpage

\end{document}